\documentclass[aps,prl,twocolumn,showpacs,preprintnumbers,amsmath,amssymb]{revtex4}

\usepackage{graphicx}
\usepackage{dcolumn}
\usepackage{bm}

\begin{document}

\preprint{LMU-ASC 22/06}
 
\title{Bulk-driven non-equilibrium phase transitions in a mesoscopic
  ring}

\author{Hauke Hinsch}
\author{Erwin Frey}%
\affiliation{%
  Arnold Sommerfeld Center for Theoretical Physics and Center of
  NanoScience,\\
  Department of Physics, Ludwig-Maximilians-Universit\"at M\"unchen, \\
  Theresienstrasse 37, D-80333 M\"unchen, Germany
}%


\begin{abstract}
  We study a periodic one-dimensional exclusion process composed of a
  driven and a diffusive part. In a mesoscopic limit where both
  dynamics compete we identify bulk-driven phase transitions. We
  employ mean-field theory complemented by Monte-Carlo simulations to
  characterize the emerging non-equilibrium steady states. Monte-Carlo
  simulations reveal interesting correlation effects that we explain
  phenomenologically.
\end{abstract}

\pacs{05.40.-a,05.60.-k, 83.50.Ha, 87.16.Nn}

\maketitle

One-dimensional (1D) driven diffusive systems \cite{schmittmann95} have not
only served as fruitful testing grounds for fundamental questions in
non-equilibrium physics \cite{derrida98} but have also been the focus
of recent interest in applications relevant to biologically problems
\cite{lipowsky01,parmeggiani}.  Restricted 1D motion may
either result from geometric confinement as in a nuclear pore complex
of cells \cite{kosztin04} or artificial crystalline zeolitical
structures \cite{kaerger}, or arise because molecular engines move
along one-dimensional tracks as for example in intracellular transport
\cite{howard} and protein synthesis \cite{macdonald68,chou03}.

In these systems interesting collective effects emerge since mutual
passage of particles is excluded. The nature of these effects depends
on whether the system is purely diffusive ({\em passive}) or driven
due to the presence of an external field or an internal driving
mechanism inherent to the particles ({\em active}), e.g.\ motor
activity in intracellular transport. In addition, one has to
distinguish between open and closed boundary conditions.  While
independent of the boundary conditions both active and passive systems
show interesting dynamic anomalies
\cite{richards77,kollmann03,bechinger}, only active systems with open
boundaries are known to exhibit non-trivial nonequilibrium
steady states \cite{krug}.

Inspired by traffic of molecular motors in closed compartments
\cite{lipowsky01} and colloidal motion in optical traps
\cite{bechinger}, we present a model that combines the symmetric (SEP)
and totally asymmetric simple exclusion process (TASEP). It is
intended to investigate
the competition between driven and diffusive motion in 1D systems.
While the breaking of translational invariance is known to be
necessary for non-trivial steady states to evolve, we show that
additionally time scale separation between the two processes requires
a mesoscopic scaling to guarantee a finite current and a physical
behavior in the continuum limit.

SEP and TASEP serve as the two paradigms for passive and active transport in
one dimension. In these lattice gas models particles occupy the sites
of a one-dimensional lattice subject to the simple exclusion rule
that each site may be occupied by at most one particle. In the SEP
particles jump independently and randomly at rate $D$ to vacant
neighboring sites with equal probabilities to the left and right,
while hopping at a rate $R$ is strictly unidirectional for the TASEP\@.
In the following we will measure time in units of $1/R$, i.e. we
set $R=1$.  For a closed ring geometry both processes are
characterized by a steady state with a uniform density profile since
translational invariance gives equal weight to all permissible
configurations. For open boundary conditions, the steady state of the
SEP always shows a linear density profile \cite{schuetz}, whose slope
depends on the boundary condition's difference.  In stark contrast,
TASEP exhibits several distinct nonequilibrium phases \cite{krug} as
a function of the magnitude of the entrance and exit rate, $\alpha$
and $\beta$, at the left and right boundary, respectively. Two
different phases can be characterized by their global particle density
(low and high) and one by a maximal current. The validity of the
initial mean-field analysis of the phase diagram was later supported
by several exact solutions \cite{derrida}.

\begin{figure}[htbp]
\includegraphics[width=8cm]{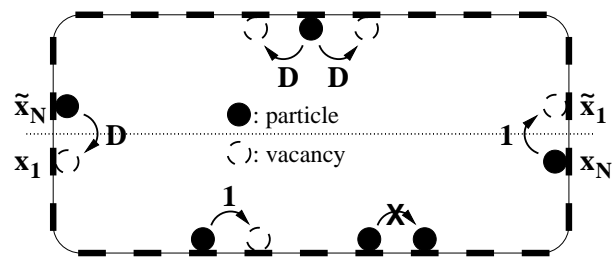}
\caption{Schematic model of the ring system. The dynamics of the 
  upper (passive) lane with lattice sites $\tilde x_i$ is governed by a SEP
  with rate $D$, and the dynamics of the lower (active) lane with
  lattice sites $x_i$ by a TASEP with unity jump rate.}
\label{fig:ring}
\end{figure}

In this Letter we aim at identifying the nature of the nonequilibrium
steady states of a closed ring system consisting of two equally sized
compartments with $i=1,\cdots,N$ lattice sites, whose dynamics is
governed by a symmetric and a totally asymmetric exclusion process,
respectively; see Fig.\ref{fig:ring}. 
To distinguish between the active and passive compartments of the
ring, quantities for the passive part like the location of the lattice
sites, $\tilde x_i$, or the occupation numbers $\tilde n_i \in \{ 0,1
\}$ are indicated by a tilde. Particle exchange between both
sub-lattices is exclusively allowed at their junctions with dynamic
rules defined by the originating site.  Thus a vacancy on the right
hand side $\tilde x_1$ of the passive part can be filled with a
particle from site $x_N$ of the active section with rate unity,
while the corresponding event at the left junction will occur with
rate $D$.  Since the system is closed, the particle number
$N_\text{p}=\sum_{i=1}^{N} (n_i+\tilde n_i)$ is conserved and the particle
density
\begin{equation}
n_\text{p}=\frac{N_\text{p}}{2 N}
\end{equation}
serves as a dimensionless control parameter confined to the interval $[0,1]$.
Having chosen the inverse hopping rate $1/R$ on the active part as our
time unit, we are left with a two-dimensional parameter space $(n_p,
D)$. As these parameters are bulk quantities, resulting phase
transitions will be bulk-induced in contrast to boundary-induced phase
transitions in TASEP\@. In order to explore the phase
behavior in this parameter space we will exploit the fact that for a
closed system the steady state current $J(n_p, D)$ is spatially
constant.

We analyze the ring system by mean-field (MF) theory complemented by
Monte Carlo simulations (MCS). The MF analysis allows to decouple the
two parts of the ring and consider them as separate lanes with
effective entrance and exit rates. Once these rates are identified we
can use known results for the SEP and TASEP with open boundaries.  To
begin with, we introduce the following notation for the stationary
densities $\rho_i=\langle n_i \rangle$ at the junction sites:
$\gamma=\tilde \rho_1$, $\delta=\tilde \rho_N$ for the passive part,
and $\alpha=\rho_1$ and $1-\beta=\rho_N$ according to TASEP convention
for the active part. The incoming current to site $x_1$ is the product
of the average occupation number on the originating site, the jump
rate to the destination site and the probability that the latter is
empty: $\delta D (1-\alpha)$. Due to current conservation this has to
equal the current to site $x_2$: $\alpha (1-\alpha)$. Using
particle-hole symmetry, one can proceed accordingly at the other
junction to arrive at the following relations:
\begin{equation}
\label{eq:interfaces}
\alpha=D \delta, \qquad \gamma=1-\beta \;.
\end{equation}

To analyze the interplay of the two junctions, we will exploit the
conservation of current mentioned above. On the active part, directed
motion results in a current of $J=\rho_i(1-\rho_{i+1})$. Conservation
of current can only be fulfilled for a spatially constant or piecewise
constant density distribution. In the latter case two sections of
constant density $U$ and $V$ are connected by a domain wall and their
densities have to fulfill the condition $\rho_U=1-\rho_V$ to conserve
the current.

As particle motion is bi-directional on the passive part, the current
between two sites is obtained as the balance of their bilateral
particle exchange which is proportional to their density difference.
Conservation of current thus demands a linear density slope
and the current takes the form $J=(\gamma-\delta)D/N$ reminiscent of
Fick's law \footnote{This might appear contradictory to the known
  non-Fickian behavior of SEP, but is due to the gradient established
  by the active part. It causes a bias that invalidates the
  sub-diffusive mean-square displacement \cite{majumdar}.}. Evidently
the passive part current vanishes with system size while the active
part current is constant. Due to current conservation this implies
that the smaller current and thus the diffusive process is dominating
the system in a trivial way. To broaden our analysis to a wide
parameter range for arbitrary system sizes we introduce a mesoscopic
scaling \cite{parmeggiani} and a new control parameter:
\begin{equation} \label{eq:scaling}
d=\frac{D}{N}   \;.
\end{equation}
This scaling can be understood as a time scale separation and ensures
competitive behavior between the system's constituents. The specific
form (\ref{eq:scaling}) even guarantees a well-behaved approach to the
continuous case $N \to \infty$.
With the new control parameter the passive part current is expressed
as $J=d (\gamma-\delta)$ and allows for a relation between the two
junctions:
\begin{equation}
\label{eq:current}
\delta=\gamma-\frac{J}{d}  \;.
\end{equation}

\begin{figure}
\includegraphics[width=9cm]{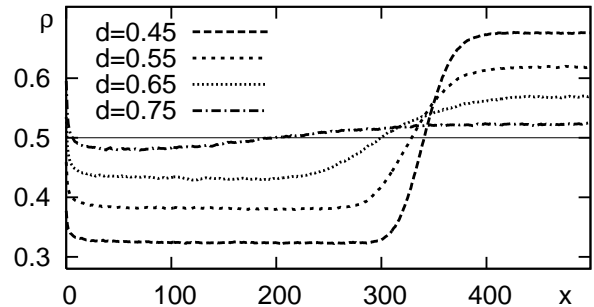}
\caption{Simulated density profiles of the active part in a system of
  $N=500$ and $n_\text{p}=0.38$ for different diffusion strengths $d$
  (indicated in the graph). The system is in the LD-HD phase featuring
  a domain wall. With increasing $d$ the domain wall position
  (intersection with the solid line) is shifted to the left (see
  Eq.~(\ref{eq:ldhd})) and DW fluctuations increase. The left boundary
  layers are due to current limitation caused by correlations at the
  passive part boundaries (see text).}
\label{fig:density}
\end{figure}

Having derived suitable entry and exit rates for the active part, we
can now apply the TASEP results. The low (high) density phases (LD,
HD) are realized in the periodic system equally and are characterized by a
uniform density below (above) $1/2$ and a boundary layer at the right
(left). Special attention has to be paid to the phase boundary
$\alpha=\beta$ between LD and HD phase, where the boundaries are
matched by a piecewise constant density profile with an intervening
domain wall (DW).  Due to the randomness of entry and exit events this
DW is delocalized and subjected to a random walk that explores the
complete system on long time scales. In the ring system however,
our MCS reveal a DW localization (Fig.\ref{fig:density}).  
We understand this as a consequence of entry and exit rates that are
not statistically independent as in TASEP but rather connected via the
passive part, dependent on the amount of available particles
and the diffusion rate.  

\begin{figure}
\includegraphics[width=8.5cm]{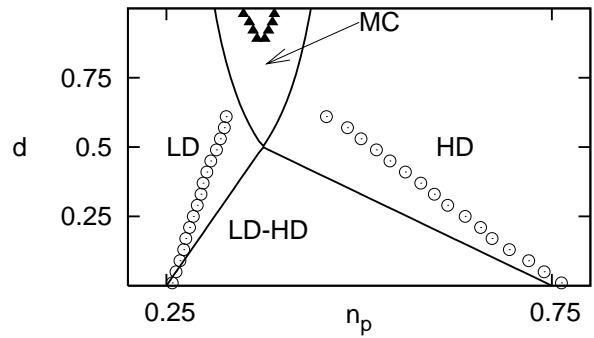}
\caption{Phase diagram obtained by MF (solid lines) exhibits
  four phases.  Simulations (MC (triangles) and LD-HD (circles) phase
  boundaries, $N=200$) reveal a failure of the MF analysis:
  correlations at the passive part boundaries cause a diminished
  current that shifts the MC phase.}
\label{fig:phasediagram}
\end{figure}

To proceed with a quantitative analysis, we connect the boundary
conditions of the two sub-lattices in a self-consistent manner to
obtain a junction density solely dependent on the two control
parameters. The precondition $\alpha=\beta$ guarantees a conserved
active part current of $J=\alpha(1-\alpha)$. This allows to rewrite
Eq.~(\ref{eq:current}) as $\delta=(1-\alpha)[1-\alpha/d]$. Since
$\delta$ is of order $1/N$ in the TL, we find $\alpha=d+d \text{
  O}(1/N)$.  Having derived $\alpha$, the remaining densities $\beta$,
$\gamma$ and $\delta$ are easily calculated and we can proceed to
connect the position of the DW to the global particle density. To this
end continuous density distributions are assumed which are applicable
for large systems where the lattice spacing vanishes. To begin with, a
DW is assumed to be present on the active part. Hence, a Heaviside
function that connects two regions of constant density at DW position
$x_\text{w}$ is chosen as
$\rho(x)=\alpha+\Theta(x-x_\text{w})(1-\alpha-\beta)$. For the passive
part's density distribution a linear slope $\tilde
\rho(x)=\delta+(\gamma-\delta)x$ is appropriate. Now particle
conservation can be expressed as $2 n_\text{p}=\int^1_0 dx [\tilde
\rho(x)+\rho(x)]$.  Solving this in the TL the DW position evaluates
to
\begin{equation} \label{eq:ldhd}
x_\text{w}=\frac{-3 + 3 d + 4 n_\text{p}}{4 d - 2} \;.
\end{equation}
The DW position is left to depend on the two control parameters: the
diffusion strength (see Fig.~\ref{fig:density} for an example) and the
particle density. The resulting ($d,n_\text{p}$)-phase diagram
exhibits several phase transitions and regimes of which three can be
characterized by the domain wall position. In the case $x_\text{w}
\leq 0$ the DW has left the active part at the left junction resulting
in an high density phase (HD) with constant density $\rho(x)=1-\beta$.
A constant low density (LD) corresponding to the left boundary
condition $\rho(x)=\alpha$ is established for $1 \leq x_\text{w}$,
while the DW is localized inside the system for $0 < x_\text{w} < 1$
connecting the two boundaries by a phase of coexistence (LD-HD).  Its
phase boundaries can be obtained from Eq.~(\ref{eq:ldhd}) as $d =
(3-4n_\text{p})/3$ for $x_\text{w}=0$ and $d = 4 n_\text{p} -1$ for
$x_\text{w}=1$ .  The phase boundaries intersect at a critical point
at $(n_\text{p}=3/8,d=1/2)$ where the DW height vanishes
(Fig.~\ref{fig:phasediagram}).
The extremal points at $d=0$ of the LD-HD phase can be readily
explained by the particle fraction of a quarter that has to be bound
on the passive part for vanishing diffusion (signifies
$\alpha=\beta=0$). Hence, at $n_\text{p}=1/4$ there are no particles
available to the active part and the system crosses over into the LD
phase. Analogous, the HD phase is entered for values of
$n_\text{p}>3/4$.  The existence of the LD-HD phase is a distinctive
difference to TASEP \footnote{Additionally, the variance of the DW
  position decays as $N^{-1}$.  Contrary to the $N^{-1/2}$ decay in
  TASEP, this is caused by the additional feedback via the passive
  part.}. It replaces one discontinuous phase transition with two
continuous transitions that meet at a multi-critical point.

Similar to TASEP, the ring system also features a maximal-current
phase (MC). In this regime the active part current imposes its maximum
$J_\text{MC}=1/4$ on both lanes. MF approximations allow again to
derive the extent of the MC phase in dependence of the two control
parameters $d$ and $n_\text{p}$. To this end we have to assume a
different density distribution than above for the active part. The
corresponding distribution in TASEP is established for boundary
conditions $\alpha,\beta>1/2$ and has to be constant at $\rho=1/2$
with the exception of possible boundary layers that vanish in the
TL\@.  Hence, $N/2$ particles have to be present on the active part.
We can then deduce from particle conservation: $2 n_\text{p} -
1/2=\delta + (\gamma - \delta)/2$ where the r.h.s is just the integral
over the passive part's linear density distribution. Using the
equality of passive part and active part current allows to solve for
$\delta = 2 ( n_\text{p} - 1/4 ) - 1/(8 d)$.  The active part
constraints can be rewritten as constraints on the passive part by use
of Eq.~(\ref{eq:current}) and Eq.~(\ref{eq:interfaces}) to $\delta >
1/(2 d N)$ and $\delta < 1/2 - 1/(4 d)$.  The last three equations
constitute planes in the ($\delta,d,n_\text{p}$)-space. Computing now
the intersection of the first plane with the two inequalities, one can
finally deduce the phase boundaries of the MC phase in
($d,n_\text{p}$)-phase space in the TL as $d=1/(16 n_\text{p} - 4)$
and $d=1/(8-16 n_\text{p})$ where the latter is the boundary with the
HD phase.  Notice that the former phase boundary could already be
derived by considering only terms of order $1$ in the TL (i.e.\ with
$\delta=0$) while the HD boundary is obtained only if
$\delta=\text{O}(1/N)$ is considered. The MC phase originates at the
critical point in phase space and asymptotes for $d \to \infty$ at
values of $1/2$ and $1/4$ (see Fig.~\ref{fig:phasediagram}) as can
easily be explained. At particle densities below $n_\text{p}=1/4$
there are not enough particles available to establish a constant
density of $\rho=1/2$ on the active part. On the contrary, the system
makes a transition to the HD phase if both active and passive part are
half-filled at $n_\text{p}=1/2$.  At this point $\gamma>1/2$ implies a
violation of the MC phase requirement $\beta>1/2$.

We have complemented our MF analysis by extensive MCS\@. Their results
deviate from the phase diagram derived above (see simulation data in
Fig.~\ref{fig:phasediagram}). The key difference is that the
multi-critical point is shifted to higher values of $d$.  To
rationalize this we compare the density profiles from simulations and
MF theory.  While the density distribution on the passive part has
been assumed to be linear according to SEP results, simulated data
exhibit a distinct curvature in the profile \footnote{We have
  conducted extensive simulations to ensure that the curvature is
  preserved at long times. There is no evidence that would hint a
  decrease in convexity with neither increasing simulation time nor
  system size.}. A closer examination of the present boundary
conditions reveals that these are different from open boundaries used
for the derivation of the linear SEP density profile.  As particles
enter the passive part with a rate of unity while the internal
dynamics of the diffusion are much faster, the SEP experiences a
boundary that is reflective during a considerable amount of time. Like
in other problems, exhibiting time scale separation, as forest-fire
models \cite{drossel92} the stationary state results from competition
between a process that strives towards equilibrium (diffusion) and
another process repeatedly driving the system to nonequilibrium
(particle entrance). The rare particle entrance events provide a
particle excess at the right boundary. This density surplus then
spreads into the bulk by diffusion before the next entrance restores
the particle excess \cite{hinsch}. A time average then results in a
strictly convex density profile on the passive part. Its curvature
renders Fick's Law for the passive part current inappropriate because
the smallest gradient (here at the left interface) acts as a current
bottleneck.  The consequences for the phase diagram are obvious: since
the current is effectively reduced the MC phase is only established at
higher diffusion values (Fig.~\ref{fig:phasediagram}). The current
limitation caused by the bottleneck is also responsible for the
boundary layers on the l.h.s.\ of the active part
(Fig.\ref{fig:density}). Here global conservation of current forces
the density to decay quickly to a value conform with the system's
minimal current. Thereby the extension of the LD-HD phase to values of
$d>1/2$ can be explained.  By introducing a correction factor for the
reduced current a refined MF theory can explain the shift of the MC
phase at least qualitatively.  Recalculation of the LD-HD phase
boundaries in this realm completely fails, also due to correlation
effects on the left interface.

Our analytical and numerical studies have shown that the presented
system exhibits a rich phase behavior unexpected for periodic systems.
While boundary-induced phase transitions in exclusion processes occur
in several geometries \cite{krug, parmeggiani} and non-trivial
nonequilibrium steady states are known in periodic systems if
translational symmetry is broken by defects \cite{lebowitz},
comparable bulk-induced phase transitions have not been studied to our
knowledge so far. Crucial to this behavior is the competition between
the two processes which the scaling Eq.~(\ref{eq:scaling}) ensures for
a broad parameter range. The time scale separation between the
subprocesses is also responsible for the quantitative failure of
MF that comes as quite a surprise, since up to now similar
approximations have been known as a reliable tool to reproduce the
phase diagram of lattice gas systems with an astonishing accuracy.

For the system presented, these characteristics indicate the existence
of interesting correlation phenomena that call for analytical methods
beyond MF. The phenomenological explanation and the modified MF
expression for the current should therefore only be considered
preliminary. As MCS near the multi-critical point require considerable
computation resources, we cannot give any reliable predictions about
the exact phase topology. The exact form of the phase diagram and in
particular the behavior near the multi-critical point, remains an open
question to whose solution different techniques like Bethe ansatz
\cite{schuetz} or density matrix RG \cite{schollwoeck} may contribute.
Furthermore, possible experimental realizations \cite{bechinger} could
make this system an intriguing problem to study.

\begin{acknowledgments}
The authors thank P.Pierobon for helpful discussions.
\end{acknowledgments}

\end{document}